# Beating the bookies with their own numbers - and how the online sports betting market is rigged

Lisandro Kaunitz[1,2*], Shenjun Zhong[3*] and Javier Kreiner[4*]

1- Research Center for Advanced Science and Technology, The University of Tokyo, Tokyo, Japan.

2- Progress Technologies Inc., Tokyo, Japan.

3- Monash Biomedical Imaging Center, Monash University, Melbourne, Australia.

4- Data Science department, CargoX, Sao Paulo, Brazil

Corresponding authors: Lisandro Kaunitz / Javier Kreiner
E-mail: lisandro@progresstech.jp / javkrei@gmail.com

# Abstract

The online sports gambling industry employs teams of data analysts to build forecast models that turn the odds at sports games in their favour. While several betting strategies have been proposed to beat bookmakers, from expert prediction models and arbitrage strategies to odds bias exploitation, their returns have been inconsistent and it remains to be shown that a betting strategy can outperform the online sports betting market. We designed a strategy to beat football bookmakers with their own numbers. Instead of building a forecasting model to compete with bookmakers predictions, we exploited the probability information implicit in the odds publicly available in the marketplace to find bets with mispriced odds. Our strategy proved profitable in a 10-year historical simulation using closing odds, a 6-month historical simulation using minute to minute odds, and a 5-month period during which we staked real money with the bookmakers[1]. Our results demonstrate that the football betting market is inefficient — bookmakers can be consistently beaten across thousands of games in both simulated environments and real-life betting. We provide a detailed description of our betting experience to illustrate how the sports gambling industry compensates these market inefficiencies with discriminatory practices against successful clients.

---

[1] Code, data and models are publicly available: https://github.com/Lisandro79/BeatTheBookie

**Introduction**

*"In the midst of chaos, there is also opportunity."*

— *Sun Tzu, The Art of War*

In recent years, the emergence of web technologies, product platforms and TV broadcast rights transformed the online gambling industry into a worldwide $452 billion business[2]. Clients of the online sports betting industry dream of "beating the bookies" and, most often, find in the adrenaline and excitement of their risky gambling activities an escape from the boredom of everyday life (Lee et al. 2007; Loroz 2004; Platz and Millar 2001; Blaszczynski, McConaghy, and Frankova 1990). To maximize profit, bookmakers employ teams of data scientists to analyze decades of sports data and develop highly accurate models for predicting the outcome of sports events (Cantinotti, Ladouceur, and Jacques 2004; García, Pérez, and Rodríguez 2016). Although several strategies have been proposed to compete with bookmakers' models, from expert predictions (Forrest, Goddard, and Simmons 2005), probability models based on Power scores, Elo ratings and/or Maher-Poisson approaches (Dixon and Coles 1997; Maher 1982; Vlastakis, Dotsis, and Markellos 2008) and prediction markets (Spann and Skiera 2009) to arbitrage strategies and odds bias exploitation (Franck, Verbeek, and Nüesch 2009; Ashiya 2015; A. Constantinou and Fenton 2013; A. C. Constantinou, Fenton, and Neil 2013), to our knowledge there is no precedent in the scientific literature that they consistently outperform the market and show sustained profit over years and across football leagues around the world (Deschamps and Gergaud 2012; A. Constantinou and Fenton 2013; Kain and Logan 2014; Spann and Skiera 2009; Vlastakis, Dotsis, and Markellos 2009, 2008).

---

[2] www.statista.com

Can a betting strategy outperform the sports betting market? Although bookmakers' profitable business (along with their modelling advantage and control of odds pricing) seems to suggest the opposite, we aimed to demonstrate that bookmakers can be beaten with their own numbers. We developed a betting strategy for the football market that exploited the implicit information contained in the bookmakers' aggregate odds (Kuypers 2000; Cortis 2016; Cortis, Hales, and Bezzina 2013) to systematically take advantage of mispriced events. Our betting system differed from previous betting strategies in that, instead of trying to build a model to compete with bookmakers' forecasting expertise, we used their publicly available odds as a proxy of the true probability of a game outcome. With these proxies we searched for mispricing opportunities, i.e., games with odds offered above the estimated fair value (see glossary in Box 1). Our strategy returned sustained profits over years of simulated betting with historical data, and months of paper trading and betting with actual money. These results suggest that the football betting market is inefficient. Bookmakers, however, deploy a special set of practical rules to compensate for these inefficiencies. A few weeks after we started trading with actual money some bookmakers began to severely limit our accounts, forcing us to stop our betting strategy. We thus demonstrate that (i) bookmakers can be beaten consistently over months/years of betting with a single strategy, in both simulated environments and in real-life betting situations and (ii) the online sports betting system is rigged against successful bettors through discriminatory practices.

**Methods**

*"Who wishes to fight must first count the cost."*

— **Sun Tzu**, **The Art of War**

*Betting strategy*

For a bet to be "fair", i.e., for the expected value of a bet to be zero, the odds paid by the bookmaker must be the inverse of the underlying probability of the result. Once bookmakers build an accurate model that estimates the underlying probability of the result of a game, they offer odds that are below the fair value. The mechanism operates similarly to the roulette at the casino. For example, when a customer places a bet on red in an American roulette, there is a 18/38 chance of doubling the wager (18 green numbers, 18 red numbers, plus 0 and 00, which are green). Under these conditions, the fair value for the bet is 2.111 but the house pays only 2 and, therefore, the house pays below fair value. This is the 'tax' or commission charged by the bookmaker, in this case, for every dollar bet at the roulette, the house expects to earn $ (2/38), or 5.3c.

In order to calculate the odds that, statistically, will allow bookmakers to earn a desired percentage of the total money bet at sport games, they need accurate models to estimate the probability of each event. There are many different factors that can be incorporated into a model to predict the probability of the outcome of a football game, for instance: the results of the last *n* games for the two teams, the record of successful games at home or away for those teams, the number of goals scored and conceded by each team during the previous games, player injuries before the game and even the expected weather conditions on the day of the match (Langseth 2013; Maher 1982; Dixon and Coles 1997). If we consider the scope of these variables the task of developing accurate models to predict the outcome of thousands of games across football leagues around the world becomes an extremely complex challenge. In recent years, however, teams of professional analysts have improved the outcomes of their prediction models with increasingly sophisticated statistical analysis and large amounts of data in variety of forms (Gandar, Zuber, and Lamb 2001).

To quantify the predictive power of bookmakers' models we conducted a historical analysis of football game outcomes. We collected the historical closing odds of football games from January 2005 to June 2015 from online sports portals available on the Internet. For this analysis we used data from 32 bookmakers: 'Interwetten', 'bwin', 'bet-at-home', 'Unibet', 'Stan James', "Expekt', '10Bet', 'William Hill', 'bet365', 'Pinnacle Sports', 'DOXXbet', 'Betsafe', 'Betway', '888sport', 'Ladbrokes', 'Betclic', 'Sportingbet', 'myBet', 'Betsson', '188BET', 'Jetbull', 'Paddy Power', 'Tipico', 'Coral', 'SBOBET', 'BetVictor', '12BET', 'Titanbet', 'Youwin', 'ComeOn', 'Betadonis', and 'Betfair'. In total, we analyzed 479,440 games from 818 leagues and divisions across the world. To measure the accuracy of bookmakers in estimating a game's final result probabilities we calculated the consensus probability as follows:

$$p_{cons} = 1/ mean(\Omega) \qquad (Eq.\ 1)$$

where $\Omega$ is a set containing the odds across bookmakers for a given event and a given game result (home team win, draw, away team win), whenever there were more than 3 odds available for that result (in some games only a subset of bookmakers offered odds; the number of odds we employed for analysis varied from a minimum of 3 to a maximum of 32). In this way, we calculated the consensus probability of a home team win, a draw, or an away team win for each of the 479,440 games (in total 3 x 479,440 consensus probabilities). Then, we binned the data according to the consensus probability from 0 to 1 in steps of 0.0125 (i.e., 80 bins). Within each bin we calculated: 1) the mean consensus probabilities across games at closing time (the final odds provided by bookmakers before the start of the match); and 2) the mean accuracy in the prediction of the football game result (i.e., the proportion of games ending in home team victory, draw or away team victory for that bin; see Figure 1). We used a minimum of 100 games for each bin. With these data we ran a preliminary analysis and observed that the consensus probability is a good predictor of the underlying probability of an outcome (see Results section). Based on these results, we decided to build our betting strategy on this evidence that bookmakers already possess highly accurate models to predict the results of football games.

A strategy intended to beat the bookmakers at predicting the outcome of sports games requires a more accurate model than the ones bookmakers have developed over many years of data collection and analysis. Instead of trying to create such a model, we decided to use the bookmakers' own probability estimates of the outcomes to find mispricing opportunities. More specifically, we searched for opportunities where some odds offered were above their estimated fair value. Sometimes bookmakers offer odds above fair value either to compete to attract clients or to maintain a balanced book to avoid getting overly exposed to risk. For example, when too many clients bet on an outcome ( e.g. home team victory) bookmakers can increase the odds for the corresponding counterpart ( e.g. away team victory), in order to attract more gamblers to bet on it and decrease their exposure to the overbooked outcome. This means that bookmakers might offer odds with a lower implied probability than the actual probability of a result. This is the key factor that we exploited in our strategy.

We based our strategy on the estimated payoff of each bet. The expected payoff of betting $1 is:

$$E(\Pi) = p_{real}(\text{outcome materializes}) \times payoff(\text{outcome materializes})$$

$$+ P_{real}(\text{outcome does not materialize}) \times payoff(\text{outcome does not materialize})$$

(Eq. 2)

$$E(\Pi) = p_{real} \times (\omega - 1) + (1 - p_{real}) \times (-1) = p_{real} \times \omega - 1 \quad \text{(Eq. 3)}$$

Where $\Pi$ is the payoff of the bet (a random variable), $p_{real}$ is the actual underlying probability that the outcome materializes, and $\omega$ are the odds paid by the bookmaker in case that the outcome comes about.

We performed a preliminary data analysis and found that:

$$p_{real} \simeq p_{cons} - \alpha \quad \text{(Eq. 4)}$$

Where $p_{cons}$ is the consensus probability as calculated above and $\alpha$ is an adjustment term that allows us to include the intercept we estimated in a regression analysis on outcomes of games for "Home", "Draw" and "Away". The estimated $\alpha$ was 0.034, 0.057 and 0.037 for home victory, draw and away victory, respectively (see results section).

Then:

$$E(\Pi) \simeq (p_{cons} - \alpha) \times \omega - 1 \quad \text{(Eq. 5)}$$

Under these conditions, we should place a bet when the expected payoff is greater than 0, i.e., when:

$$\omega > 1/(p_{cons} - \alpha) \quad \text{(Eq. 6)}$$

We followed this line of reasoning to define our betting strategy, and decided to place a bet whenever the maximum odds offered for a given result fulfilled the following inequality:

$$max(\Omega) > 1/(p_{cons} - 0.05) \quad \text{(Eq. 7)}$$

The expected value of each bet increases with the $\alpha$ parameter, while the number of games available for betting decreases. This occurs because the condition becomes more stringent and less bookmakers offer odds with such high margins. To select an appropriate value for the $\alpha$ parameter we analyzed the performance of the simulation strategy by varying the value of $\alpha$ from 0.01 to 0.1. We found that an $\alpha$ of 0.05 produced the optimal payoff with the largest amount of games (an $\alpha$ of 0.06, for example, was equally profitable but we decided to use an $\alpha$ of 0.05 because it provided twice as many games to bet in, which might be useful in a strategy that increases the amount staked as a function of the earnings).

In summary, we based our betting strategy on the assumption that odds published by bookmakers allow us to obtain a highly accurate estimate of the actual probability of the outcome of an event (by taking the inverse of the mean odds across bookmakers minus a constant that allows for the bookmaker's commission). Thus, our betting strategy consisted of placing bets whenever the odds offered by some bookmaker deviated from the average and were above fair value, i.e., when the expected payoff of placing the bet was positive. Importantly, the task of identifying the odds that satisfied the threshold in (Eq. 7) did not require a model with higher accuracy than the bookmakers' models.

*Strategy implementation*

Our betting strategy was implemented as a real time system, and deployed on a virtual machine hosted on the cloud. The system continuously collected data from online sports betting portals and provided the web service that made a dashboard available, where the recommended bets and the betting history were shown (Supplementary Figure 1). The virtual machine was used to run a program that searched for the odds of every football game from 5 hours before the onset of the game. For each game, the program continuously collected odds across 32 bookmakers and calculated whether the maximum offered odds complied with our strategy's condition for placing a bet - i.e., maximum odds fulfilling, Eq. (7). Whenever the program found a situation in which this happened, it displayed the information about the game, bookmaker and odds on the dashboard (Supplementary Figure 1), so that the users (including us) could see the list of bets recommended by the system and place a bet of fixed amount with the bookmaker. To keep the amount of money placed on each independent game constant, once a bet was placed for a game at some bookmaker, that game was not considered for further analysis.

**Results**

*"Victorious warriors win first and then go to war …. The greatest victory is that which requires no battle."*
*— **Sun Tzu**, **The Art of War***

*Analysis to define the betting strategy*

To select the appropriate strategy we first performed a descriptive statistical analysis of the relationship between the bookmakers' predictions and the actual probability of the outcome of football games. A linear regression analysis showed a strong correlation between the bookmakers' consensus probability and the results of the game for home victory ($R^2 = 0.999$), draw ($R^2 = 0.995$) and away victory ($R^2 = 0.998$). The slopes and intercepts of the regression line were 1.003 and -0.034 for a home victory, 1.081 and -0.057 for draw, and 1.012 and -0.037 for an away victory, respectively. These results suggest that the consensus probability is an extremely accurate proxy (up to a constant intercept) of the actual probability of occurrence of each event (home victory, dray, away victory; note that the slopes of the three regression lines are very close to 1). Based on these results, we decided to build our betting strategy on the evidence that bookmakers already possess highly accurate models to predict the results of football games.

*Strategy Outcome*

We tested our betting strategy by analyzing the odds and results of 479,440 football games played in 818 leagues during a ten-year period, from 2005 to 2015. We began our analysis by applying our betting strategy to the closing odds of each game (i.e., the odds values offered by bookmakers at the start of the game[3]). We simulated placing bets when the closing odds of a bookmaker complied with Eq. (7) at the closing time of the odds. With this approach, our betting strategy reached an accuracy of 44.4% and yielded a 3.5% return over the analysis period. For example, for an imaginary stake of $50 per bet, this corresponds to an equivalent profit of $98,865 across 56,435 bets (Table 1, Figure 2A). We performed a bootstrap analysis to assess whether our returns were above chance level. We repeatedly simulated a strategy that chooses a random sample of games and, for each game in the sample, randomly selects to bet for home, draw or away (with a prior probability equal to that of our strategy, see below) and places wagers at the maximum odds offered across bookmakers. In each run of the simulation we A) randomly sampled 56,435 games (the same amount of games that were selected by our betting strategy) from the complete set of games in the historical series, B) selected a random outcome on which to bet with a probability of 0.595 for home victory, 0.021 for draw and 0.384 for away victory (these are the proportions of home, draw and away games that were selected by our strategy) and C) calculated the return of placing the bet. We repeated the procedure 2000 times (sampling with replacement) to obtain a distribution of returns (Figure 2A). The random strategy yielded an accuracy of 38.9%, an average return of -3.32% and an average loss of $93,563 (STD=$17,778), further confirming that Eq. 7 successfully selects bets with a positive expected payoff above chance level. The return of our strategy was 10.82 standard deviations above the mean return of the random bet strategy. The probability of obtaining a return greater than or equal to $98,865 in 56,435 bets using a random bet strategy is less than 1 in a billion.

---

[3] In practice, closing odds are a particular case of odds because they reflect the latest odds that were available for clients to bet. We note, however, that these odds values are not a perfect estimate of a system that could be used in real life because bookmakers close their odds shortly before the start of the game.

We observed that the final accuracy was higher for our strategy (44.4%) than for the random bet strategy (38.9%). Correspondingly, our strategy selected odds with a mean value of 2.30 (STD=0.99) and the random bet strategy selected odds with a mean value of 3.10 (STD= 2.42). The discrepancy in the accuracy between strategies originated from the selection of events: our strategy picked up games with lower odds values and higher probability of occurrence than the games selected by the random bet strategy. We confirmed this finding with an analysis of the mean closing odds across bookmakers for each strategy. As shown above, the mean closing odds across bookmakers is a precise estimate of the true probability of occurrence of an event (Figure 1). The expected accuracy (as predicted by the inverse of the mean closing odds across bookmakers) precisely estimates the final accuracy in each strategy. We calculated the expected accuracy of the strategy using Eq. 4. For each bet of the strategy we calculate

$$p^i_{cons} = 1/mean(\Omega) - \alpha \qquad \text{(Eq. 8)}$$

where α is equal to 0.034, 0.057 and 0.037 for home win, draw and away bets respectively (and where the intercept α comes from the regression analysis performed in the first paragraph of the Results section), and $i$ indexes the bet. We then calculated the average estimated probability: $E(accuracy) \simeq mean(p^i_{cons})$. The expected accuracy for our strategy was 45.9%, and the actual accuracy was 44.4%, while the expected accuracy for the random bet strategy was 38.9% and its actual accuracy 38.9%. Although the mean closing odds values differed between strategies, the final accuracies of both strategies closely matched the expected accuracy according to Eq. 8. This confirms that the the probability information implicit in the mean closing odds across bookmakers represents a powerful predictor for the true outcome of football games (as shown in our historical analysis).

Following the success of our initial analysis, and considering that in real life individuals cannot place bets at the closing time of odds, we decided to conduct a more realistic simulation in which we placed bets at odds available from 1 to 5 hours before the beginning of each game. To this end, we wrote scripts to continuously collect odds from multiple sources on the Internet. While the historical closing odds for football games can be easily retrieved online, we could not find any source of data containing the time series of odds movements before the beginning of each game. To obtain these times series we wrote a new set of scripts to gather information in real time for upcoming games as they became available online. In total, we were able to obtain data from 31,074 games, from the 1st of September 2015 to the 29th of February 2016. Using these times series data, we placed bets according to our betting strategy at any time starting 5 hours before the beginning of a game until 1 hour before the start of the game. Under these simulated conditions, our strategy selected odds with a mean value of 2.32 (STD=0.99.), had an accuracy of 47.6% and yielded a 9.9% return; i.e., if every bet placed was $50 our strategy would have generated $34,932 in profit across 6,994 bets (Table 1, Figure 2B). In contrast, the distribution of returns of the random bet strategy selected odds with a mean value of 3.29 (STD=2.96), had and accuracy of 38.4% and would have generated, for bets of $50, a return of 0.2% and an average profit of $825 (STD=$7,106). Similarly as shown above, the expected accuracy wa 46.5% for our strategy and 37.7% for the random bet strategy, which closely matched the actual accuracies of both strategies. The return of our strategy was 4.80 standard deviations above the mean of the random bet strategy. The probability of obtaining a profit greater than or equal to $34,932 in 6,994 bets with a random bet strategy is less than 1 in a million.

Once we determined that our betting strategy was successful with the historical closing odds and with the analysis of odds series movements from 5 hours to 1 hour before the game start, we decided to test our betting strategy under more realistic betting conditions. To this end we employed a technique called "paper trading", a simulated trading process in which bettors can "practice" placing bets without committing real money. We used the information displayed on the dashboard to check the bookmakers' accounts, verify that the possibility to lay a bet at the advantageous odds was available, and subsequently mark the bet as laid on the dashboard. Paper trading allowed us to empirically check whether the odds were available at the bookmakers at the time of placing a bet. We had to test the discrepancy between the information that bookmakers showed on their websites and the information that was displayed on our dashboard. Often, there was a time delay between the moment when bookmakers made their odds available online and the time it took for our scripts to show that information on the dashboard. We observed that around 30% of the odds that were displayed on the dashboard had already been changed at the bookmakers' sites. The delay in the update of the odds created a sample bias in the games we were betting on: in contrast to previous analysis in which every game was used for the simulation, now a subset of these games was not included at the time of placing bets. To test how this delay could affect our results, we ran again our strategy simulation, now randomly discarding 30% of the games. We observed that, despite the missing bets, the strategy remained profitable. We decided to continue with our betting strategy, and after three months of paper trading our strategy obtained an accuracy of 44.4% and a return of 5.5%, earning $1,128.50 across 407 bets for the case of $50 bets (Table 1, Figure 3).

At this point we decided to place bets with real money. All the procedures were identical to the paper trading exercise, with the exception that the human operator actually placed $50 bets at the bookmakers' online platforms after checking the odds on the dashboard. Our final results show the profit we obtained in 5 months of betting money for real. During that period we obtained an accuracy of 47.% and a profit of $957.50 across 265 bets, equivalent to a 8.5% return (Table 1, Figure3). Combined, paper trading and real betting had an accuracy of 45.5% and yielded a profit of $2,086 in 672 bets, equivalent to a return of 6.2%. We compared the results of our strategy with the results of a random bet strategy, identical to that employed for the time series odds (figure 2B) but this time considering games from April 2015 to July 2015 (the period used for paper trading and real betting). The random strategy yielded an accuracy of 38.7%, an average return of -0.7% and an average loss of $670 (STD=$2047). The return of our strategy after 672 games was 1.34 standard deviations above the mean of the random bet strategy and the probability of obtaining a profit of $2,086 or higher in 672 bets with a random betting strategy is 1 in 11. This probability corresponds to a p value of 0.089, under the null hypothesis that the return of our strategy comes from a distribution of final returns obtained with a random bet strategy. A p-value of 0.05 is often considered as the standard threshold for statistical significance. The p-value we obtained from the analysis of the return of our strategy was expected given the evolution of the returns obtained in our historical simulations: with an increase in the number of games our strategy increases its return and separation from the distribution of returns of the random bet strategy (as seen with the historical analysis of closing odds and odds movements series). The reader might notice that during a similar time period the simulated strategy bet on approximately ten times more games. The reason for this is that we did not have a dedicated operator betting on all available opportunities 24 hours a day and as a result we missed many of the bets that appeared on the dashboard. Nevertheless, our paper trading and actual betting activity confirmed the profitability of the strategy.

Although we played according to the sports betting industry rules, a few months after we began to place bets with actual money bookmakers started to severely limit our accounts. We had some of our bets limited in the stake amount we could lay and bookmakers sometimes required "manual inspection" of our wagers before accepting them. In most cases, bookmakers denied us the opportunity to bet or suggested a value lower than our fixed bet of $50 (Figure 4). Under these circumstances we could not continue with our betting strategy. The limits imposed by bookmakers not only shrunk our potential profit but also created a sampling bias in the choice of games which was not taken into account in our previous analysis. In our simulations, when we analyzed the effects of randomly discarding a proportion of the games, the returns were not affected. However, the selection of games where bookmakers limited our stakes was unlikely to be purely random, which could negatively impact the strategy's performance. For these reasons,

and because bookmakers' restrictions turned the betting experience increasingly difficult, we decided to end our betting experiment[4].

**Discussion**

*"One may know how to conquer without being able to do it."*

— ***Sun Tzu**, The Art of War*

We developed a betting strategy for the online betting football market. In contrast to strategies that build prediction models to compete with the forecasts of bookmakers' models, our strategy was developed under the assumption that the average of the odds across bookmakers reflects an accurate estimate of the probability of the outcome of a game. Instead of competing against bookmakers' forecasting models, we used the prediction information contained in the aggregate odds to bet on mispriced events. Our strategy proved successful and returned profit with historical data, paper trading and real betting over months and across football leagues around the world.

Betting strategies based on expert or tipster analysis attempt to beat bookmakers by constructing more accurate forecasting models than those of bookmakers (Deschamps and Gergaud 2012; A. Constantinou and Fenton 2013; Vlastakis, Dotsis, and Markellos 2009; Daunhawer, Schoch, and Kosub 2017; Boulier, Stekler, and Amundson 2006). Our analysis shows, however, that the implicit information contained in the average odds across bookmakers provides a highly accurate model to predict the outcomes of football games (Forrest, Goddard, and Simmons 2005; Boulier, Stekler, and Amundson 2006; Spann and Skiera 2003). There are many cases where the aggregate predictions of a group of individuals produce more accurate predictions than those of each individual separately, a phenomenon often referred to as *the wisdom of crowds* (Navajas et al. 2017). This idea is often applied in practice, for example in applications such as ensemble learning in machine learning algorithms (Géron 2017). Similarly, in the football market, each bookmaker can be considered a predictor, and the average odds as the aggregate information across predictors. These predictions also include the preferences and opinions of the punters regarding the probability of the outcome, because they exert pressure on the price of the odds through their collective betting (bookmakers often alter odds based on demand level to keep a balanced book, e.g. when they increase the odds for a favourite when a disproportionate amount of punters place money on the underdog). As bookmakers already posses excellent predictive models to estimate the outcomes of football games, competing with them at forecasting game outcomes becomes a challenging task. Not surprisingly, previous attempts to beat the football market with expert strategies showed inconsistent returns (Deschamps and Gergaud 2012; A. Constantinou and Fenton 2013; Kain and Logan 2014; Vlastakis, Dotsis, and Markellos 2009; Forrest, Goddard, and Simmons 2005; Vlastakis, Dotsis, and Markellos 2008; Daunhawer, Schoch, and Kosub 2017; Boulier, Stekler, and Amundson 2006). In comparison, our strategy showed positive and sustained returns over years of betting with historical data and over months of betting actual money across leagues in the football market.

---

[4] As of the date of writing this paper (August 2017), one of the bookmakers we had accounts with, "Doxxbet", closed its website to clients. We are not able to withdraw the money (90 euro) from them. Their support teams do not respond to our emails.

Through our experiments we demonstrated the existence of a betting strategy that consistently generates profit. Some scholars consider that the existence of one such strategy is inconsistent with the putative "efficiency" of the betting market (Deschamps and Gergaud 2012; A.

Constantinou and Fenton 2013; Vlastakis, Dotsis, and Markellos 2009). If, on the contrary, a strategy like ours generates profit consistently either by outperforming bookmakers' predictions or by exploiting market flaws then the betting market is necessarily "inefficient". Our results suggest that the online football betting market is inefficient because our strategy was able to obtain sustained profits over years with historical data and over months of paper trading and actual betting. In practice, however, the inefficiency of the football betting market was compensated by the bookmakers' restrictive practices. A few months after we began placing bets with real money bookmakers limited our accounts, which forced us to stop our betting completely. Although our betting activities were legal and were conducted according to the bookmakers' rules, our bet stakes were nevertheless restricted. Our case illustrates some of the discriminatory practices of the online sports betting market − the sports betting industry has the freedom to publicize and offer odds to their clients, but those clients are expected to lose and, if they are successful, they can be restricted from betting. In comparison, the limits to the accounts imposed in the online gambling industry constitute illegal practices in other industries, or are even unlawful in general. For example, advertising goods or services with intent not to sell them as advertised, or advertising goods or services with no intent to supply reasonably expectable demand but with the intention to lure the client to buy another product (a practice, often called "bait" or "bait and switch" advertising) is considered false advertising and carries pecuniary penalties in the UK, Australia and the United States of America[5]. Most countries have laws regulating advertising in the gambling industry, but some of these laws have been relaxed in recent years (e.g. the Gambling Act 2005 in the UK allowed the sports gambling industry to start advertising online and on TV) and they vary from country to country. Our study sets a precedent of the discriminatory practices against successful bettors in the online sports gambling industry: the online football market is rigged because bookmakers discriminate against successful clients. We advocate for governments to take action into further regulating the sports betting industry, either by forcing bookmakers to publicly admit that successful clients will be banned from betting or by denying bookmakers the chance to discriminate against them.

---

[5] *Consumer Protection from Unfair Trading Regulations (2008) Guidance, Interim: Guidance on the UK Implementation of the Unfair Commercial Practices Directive*
(https://www.gov.uk/government/uploads/system/uploads/attachment_data/file/284442/oft1008.pdf)
Australia: COMPETITION AND CONSUMER ACT 2010 - SCHEDULE 2
(http://www.austlii.edu.au/au/legis/cth/consol_act/caca2010265/sch2.html)
US: FTC Guides Against Bait Advertising, Section 238.

**Acknowledgements**

We are very grateful to Katia Giacomozzi, who performed the bulk of the task of placing actual bets and dealing with bookmakers' accounts. This work could not have been possible without her hard work and enthusiastic contribution. We are also very grateful to Ben Fulcher, Adrian Carter and Alessio Fracasso for their comments on an earlier version of the manuscript. Lisandro Kaunitz was supported by a fellowship from the Japanese Society for the Promotion of Science (JSPS P15048).

**Figures and Tables**

**Table 1.** Results obtained with historical data, paper trading conditions and real betting.

|  | **Dates** | **Profit (U$D)** | **Return** | **Number of bets** | **Accuracy** |
|---|---|---|---|---|---|
| **Historical Closing odds** | 01/01/2005 to 30/06/2015 | 98,865 | 3.5 % | 56,435 | 44.4% |
| **Continuous odds** | 01/09/2015 to 29/02/2016 | 34,932 | 9.9 % | 6,994 | 47.6% |
| **Paper trading** | 06/03/2016 to 19/04/2016 | 1,128.5 | 5.5 % | 407 | 44.5 % |
| **Actual betting** | 23/04/2016 to 18/09/2016 | 957.5 | 8.5 % | 265 | 47.2 % |

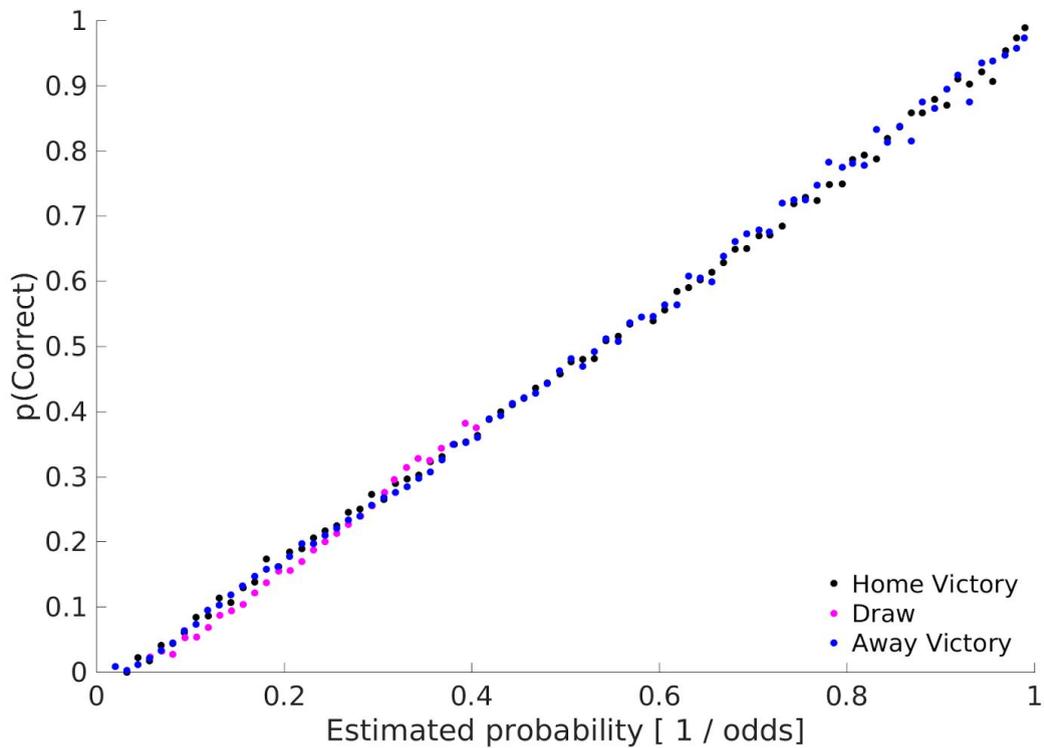

**Figure 1**. Bookmakers prediction power. A historical analysis of 10 years of football games shows the tight relationship between the bookmakers' predictions and the actual outcome of football games. The probability estimated by bookmakers (as reflected by the inverse of the mean closing odds across bookmakers) is highly correlated with the true probability of the outcomes of football games for home team victory (black dots), draw (magenta dots) and away team victory (blue dots). We analyzed 479,440 games from 818 leagues and divisions across the world during the period 2005-2015. Data was binned according to the estimated probabilities, from 0 to 1 in steps of 0.0125, and with a minimum of 100 observations per bin.

**A**

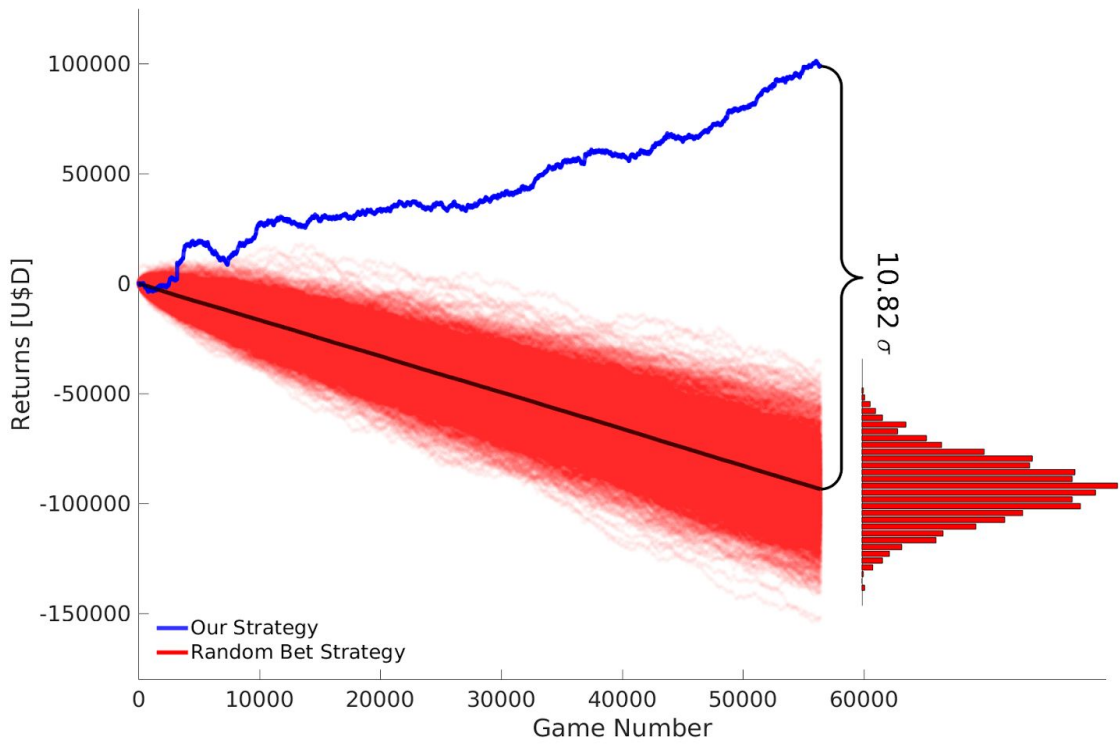

**B**

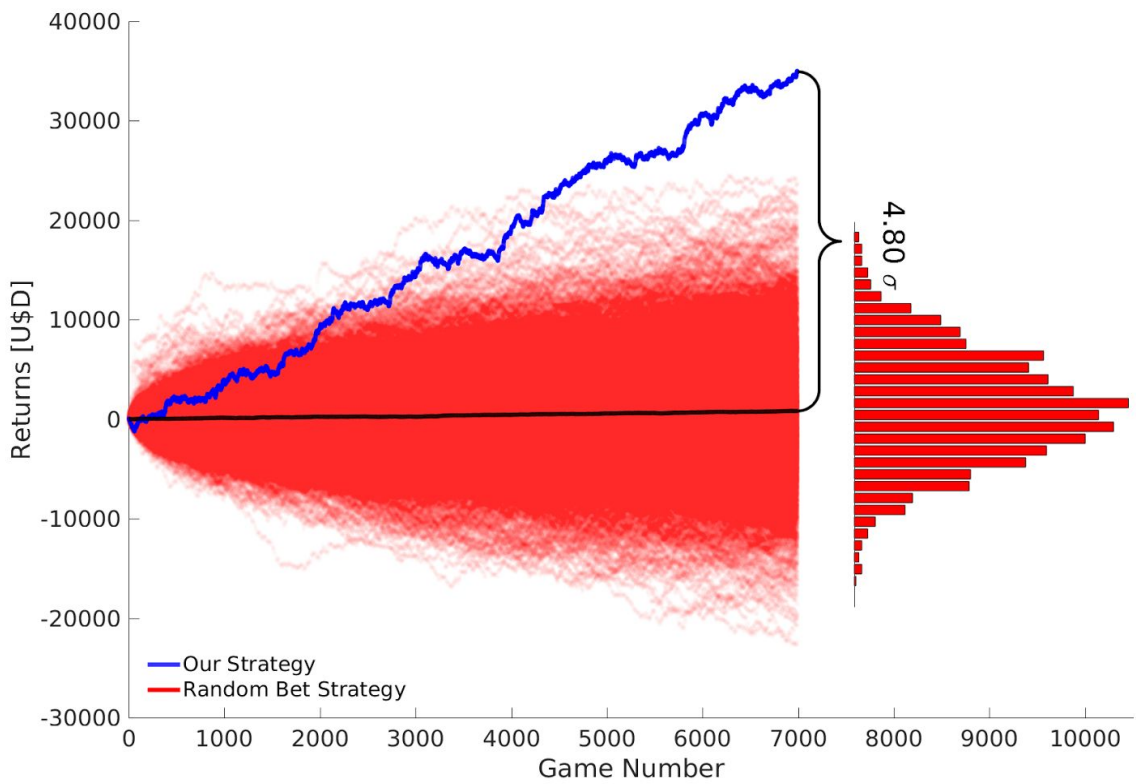

**Figures 2**. Two analysis with historical data demonstrate the effectiveness of our betting strategy. A- We applied our strategy to the closing odds of 479,440 games and obtained a return of 3.5% in 56,435 bets. To assess the probability of obtaining a return greater than or equal to 3.5% by chance we performed a bootstrap analysis to estimate the distribution of returns for a "Random Bet Strategy". By placing bets at the highest offered odds at random games the "Random bet Strategy" yielded, on average, a return of -3.32%. In comparison, the return of our strategy was 10.82 standard deviations above the mean of the distribution of returns of the random bet strategy. The probability of obtaining a return greater than or equal to ours with a random bet strategy across 56,435 games is less than 1 in a billion. Data in this panel comes from a 10-year database (2005-2015) of football games. The figure shows the potential total return assuming a constant $50 stake per bet. B) We applied the same bootstrap analysis as in A), but now to the time series of odds movements during the period [-5 -1] hours before the start of the games. The random bet strategy yielded an average return of 0.2%. In comparison, the return of our strategy was 9.9%, 4.80 standard deviations above the mean of the distribution of returns of the random bet strategy. The probability of obtaining a return greater than or equal to ours with a random strategy that bets on the maximum odds across 31,074 games is less than 1 in a million. Data in this panel comes from a 6-month database of football games (September 2015 - March 2016).

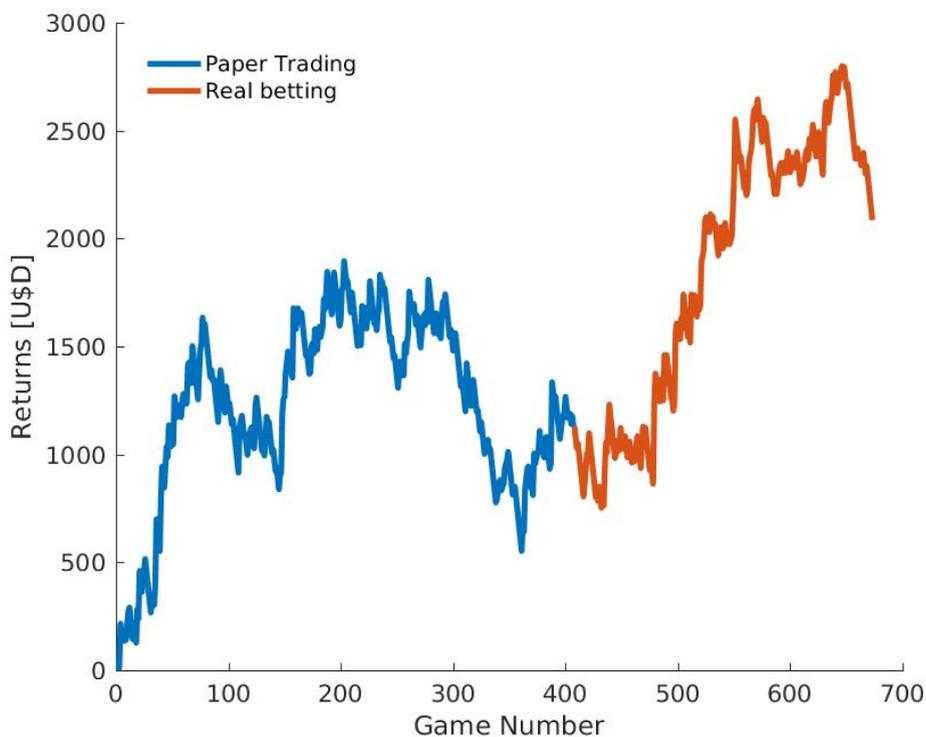

**Figure 3.** Our betting strategy generated profit with paper trading and in a real-life betting (placing real stakes with bookmakers). We obtained a return of 5.5% for "paper trading" (blue

line) and a return of 8.5% for real betting (see Table 1 for a detailed analysis) over a 5-month period of betting. Considering both paper trading and real betting we made a profit of $2,086 in 672 bets, a return of 6.2%. This was achieved by placing $50 on each bet.

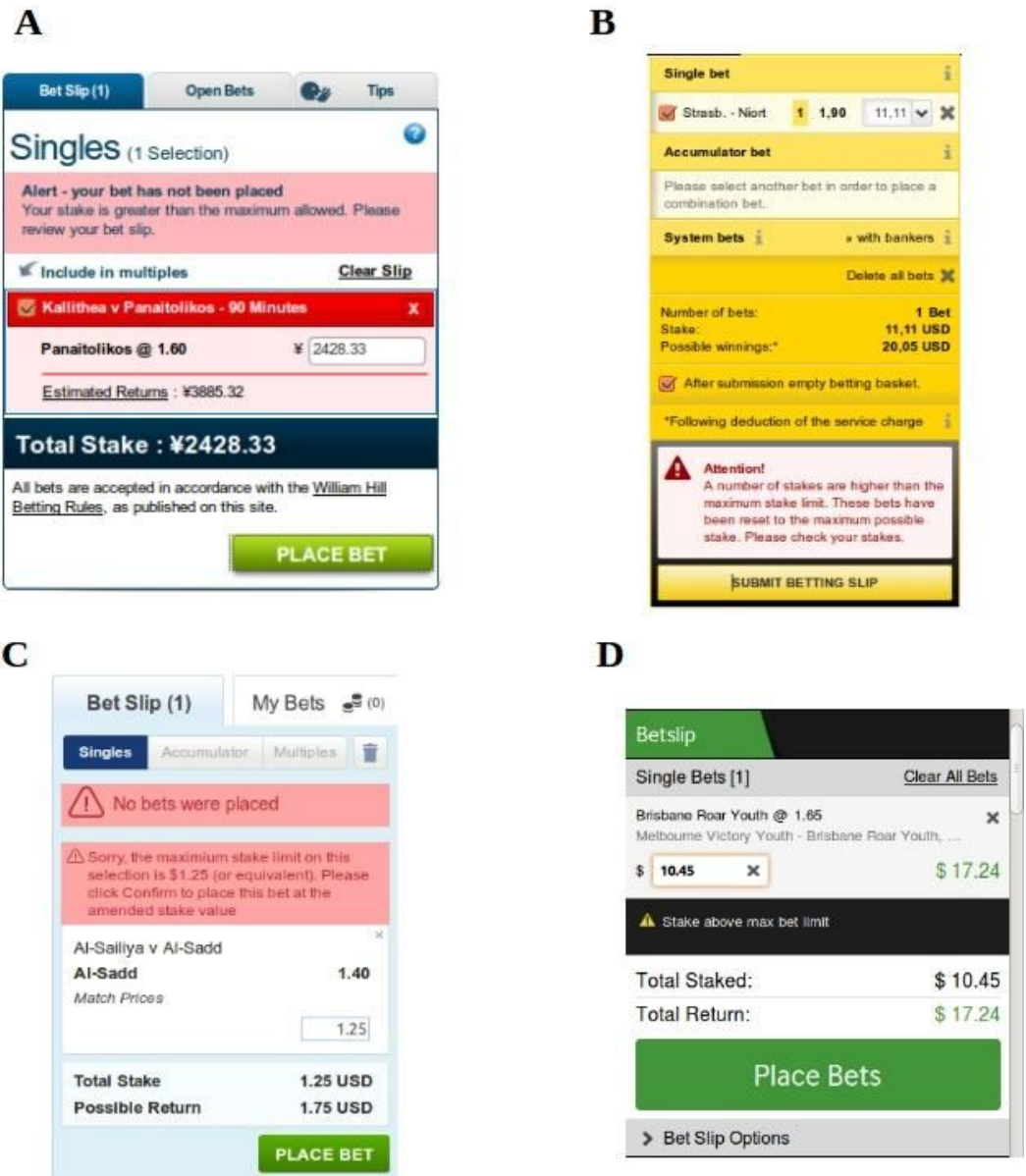

**Figure 4.** Bookmakers discriminate against successful clients. Betting limits set on some of our stakes. The figure shows examples of such limits imposed on our accounts by four Bookmakers. A- William Hill betting slip showing (www.sports.williamhill.com/bet/en-gb) a 2,428.33 yen limit on our bet (at the time this bet took place 5000 yen were equivalent to $50). B- Interwetten (www.interwetten.com) imposing a maximum bet of $11.11 C- Sportingbet (http://www.sportingbet.com/) setting a maximum limit of $1.25 and D- Betway (www.betway.com) limiting our stakes to $10.45.

**Supplementary Materials**

| | |
|---|---|
| **Box 1. Glossary** | |
| Bookmaker | The bookmaker (or "bookie"), refers to the company that provides an odds market for betting and offers to pay a price for each possible outcome of a sporting event. |
| Event | This term denotes a specific match between two teams or individuals. For example: Ath. Bilbao vs Barcelona, Thursday, January 5, 21:15 GMT. |
| Market | A betting market is a type of betting proposition with two or more possible outcomes. The result of the match (home win, away win, or draw), the number of goals scored (two or less goals, three or more), or the time of the first goal are a few examples of different markets for a single sporting event. |
| Stake | The amount of money wagered in a single bet. |
| Odds | The odds of a result refer to the payoff to be received if the chosen result materializes. In this paper we use the European notation, where the odds are equal to the currency units to be received for each currency unit wagered. For example, if an outcome offers odds of 2 means that for each dollar wagered the house will pay 2 back, giving a profit of 1 dollar per dollar invested. |
| Fair odds | Fair odds for an outcome are the ones that result in a zero expected payoff. For example, if the probability of the outcome is ½, the fair odds would be 2, because E(payoff) = ½ x (2 - 1) + ½ x (-1) = 0. In general for the odds of an outcome to be fair, they should be the inverse of the probability of the outcome. |

| **Result/Outcome** | The actual outcome of an event. E.g. in a 1x2 soccer bet, local win, draw, and away win are the three possible results. If the result that comes about coincides with the chosen result of a bet, the gambler wins the odds times the stake, otherwise he loses the whole stake. |
|---|---|
| **Profit** | The amount of additional money the bettor receives on top of his stake if he chooses the result that actually happens. For example, if the odds are 2 and the stake is 10 dollars, the gambler receives 20 dollars in total from the bookie, and the profit is 10 dollars. |
| **Yield** | A measure of the profitability of a series of bets, it is calculated as the sum of the profits made from all the bets placed divided by the sum of the money staked in all bets, usually expressed as a percentage. For example, if after 10 bets of $1 each there is a net profit of $1.50, the yield is (1.5/10) = 0.15=15%. |

A

Dashboard | Bets List

soccer|1x2 ▼                    Choose >>

| | Sport | Match Title | League | Result | Result Det | Date | Time to Match | Best Bookie | Best Odds | Mean / Median | Bet Amount |
|---|---|---|---|---|---|---|---|---|---|---|---|
| 👁 | soccer | Rabotnicki vs. Pobeda | FYR of Macedonia: First League | 1 | | 2016-11-30 13:00:00 | 00:17:40 | William Hill | 1.57 | 1.44722222 / 1.45 | Bet / Check |
| 👁 | soccer | Sileks vs. Shkupi | FYR of Macedonia: First League | X | | 2016-11-30 13:00:00 | 00:17:40 | William Hill | 4.20 | 3.46529411 / 3.35 | Bet / Check |
| 👁 | soccer | Naftovik-Ukrnafta vs. Dyn. Kiev | Ukraine: Ukrainian Cup | 2 | | 2016-11-30 16:00:00 | 03:17:40 | Ladbrokes | 1.18 | 1.10272727 / 1.11 | Bet / Check |

B

| | | Dashboard \| Bets List | | | | | | | | | | |
|---|---|---|---|---|---|---|---|---|---|---|---|---|
| | | **Total Bets:** 407 | | **Total Profit:** 1128.5 | | | **Accuracy:** 0.44 | | | **Mean odds:** 2.73 | | |
| 01/22/2016 | 04/22/2016 | Restrict >> | | | | | | | | | | |
| | Match Title | League | Bet Result | Match Result | Match Result Det | Date | Time to Match | Bet Odds | Bet Bookie | Bet Amount | Win/Loss | Profit |
| | ES Setif vs. Al-Merreikh | Africa: CAF Champions League | 1 | 0:0 | (0:0, 0:0) | 2016-04-19 20:00:00 | 2016-04-19 20:00:00 | 1.60 | Betsson | 50 | -1 | -50 |
| | Princesa Solimoes vs. Chapecoense-SC | Brazil: Copa do Brasil | 2 | 1:2 | (1:1, 0:1) | 2016-04-06 21:30:00 | 2016-04-06 21:30:00 | 1.78 | bwin | 50 | 1 | 39 |
| | Shakhtyor Soligorsk vs. FC Minsk | Belarus: Belarusian Cup | 1 | 0:0 | (0:0, 0:0) | 2016-04-06 17:30:00 | 2016-04-06 17:30:00 | 1.60 | bwin | 50 | -1 | -50 |
| | Renova vs. Carev Dvor | FYR of Macedonia: First League | 1 | 3:0 | (0:0, 3:0) | 2016-04-06 16:00:00 | 2016-04-06 16:00:00 | 1.14 | bwin | 50 | 1 | 7 |
| | Bregalnica Stip vs. Met. Skopje | FYR of Macedonia: First League | 1 | 4:1 | (2:0, 2:1) | 2016-04-06 16:00:00 | 2016-04-06 16:00:00 | 1.29 | William Hill | 50 | 1 | 14.5 |
| | Saham vs. Al Nasr | Oman: Sultan Cup | 1 | 0:0 | (0:0, 0:0) | 2016-04-06 15:45:00 | 2016-04-06 15:45:00 | 2.40 | 888sport | 50 | -1 | -50 |
| | Dobrovice vs. Chrudim | Czech Republic: CFL | 1 | 1:4 | (0:1, 1:3) | 2016-03-30 16:30:00 | 2016-03-30 16:30:00 | 2.50 | bwin | 50 | -1 | -50 |

**Supplementary Figure 1.** A) Screenshot of the online dashboard displaying the games that were selected for betting. The dashboard displayed the names of both teams, league name, bet value at which to place the bet for the strategy to work and the time remaining until the start of the game. B) A second tab in the Dashboard was used to keep track of the bets list. There the dashboard displayed the names of the participating teams, football league, the result that was backed by the bet (1: home team to win, 2: away team to win), final result of the game, odds value for the bet, the bookmaker that was used for each bet, the amount of money placed on each bet (we employed U$D 50 throughout), the result of the bet (1: bet won; -1: bet lost) and the profit obtained from each bet. Some of the games used for paper trading are displayed in this figure.